\documentclass[epj,numbook]{svjour}
\usepackage{graphics,epsfig,amsmath,amssymb,isolatin1}
\begin{document}
%\preprint{WUP ..}
\title{A numerical study of the formation of magnetisation plateaus in quasi
  one-dimensional spin-1/2 Heisenberg models}
\titlerunning{Formation of plateaus in Heisenberg models}
\authorrunning{Wie{\ss}ner} 
\author{R.M. Wie{\ss}ner, A. Fledderjohann, K.-H.
  M{\"u}tter, and M. Karbach} 
\institute{Physics Department, University of
  Wuppertal, D-42097 Wuppertal, Germany} 
\date{\today}
%%%%%%%%%%%%%%%%%%%%%%%%%%%%%%%%%%%%%%%%%%%%%%%%%
% ABSTRACT
%%%%%%%%%%%%%%%%%%%%%%%%%%%%%%%%%%%%%%%%%%%%%%%%%
\abstract{We study the magnetisation process of the one dimensional spin-1/2
  antiferromagnetic Heisenberg model with modulated couplings over $j=1,2,3$
  sites. It turns out that the evolution of magnetisation plateaus depends on
  $j$ and on the wave number $q$ of the modulation according to the rule of
  Oshikawa \textit{et al.}. A mapping of two- and three-leg zig-zag ladders on
  one dimensional systems with modulated couplings yields predictions for the
  occurence of magnetization plateaus. The latter are tested by numerical
  computations with the DMRG algorithm.}

\maketitle
%\onecolumn
%%%%%%%%%%%%%%%%%%%%%%%%%%%%%%%%%%%%%%%%%%%%%%%%%
%
\section{Introduction}
\label{sec:Introduction}
%
%%%%%%%%%%%%%%%%%%%%%%%%%%%%%%%%%%%%%%%%%%%%%%%%%
The formation of gaps and plateaus in the magnetisation process of
one-dimensional (1D) spin-1/2 antiferromagnetic Heisenberg models has been
studied intensively during the last
years~\cite{OYA97,Hida94,Tots98,CHP97,TNK98,CHP98,FKM98,CG99,FKM99,FGK+99}.

Oshikawa, Yamanaka and Affleck \cite{OYA97} pointed out the crucial role which
play the \textit{soft modes}, predicted by the Lieb-Schultz-Mattis (LSM)
construction~\cite{LSM61}. For example in the case of the 1D spin-1/2 Hamiltonian
with nearest neighbour couplings and a homogeneous external field $B$:
\begin{eqnarray}
  \label{eq:1}
  {\bf H}(B) &\equiv& {\bf H}_{1}-B{\bf S}_3(0), \\
  {\bf H}_{j} &\equiv& 2\sum_{l=1}^{N}{\bf S}_l \cdot {\bf S}_{l+j}, \quad
  n=1,2,\ldots, \\
  {\bf S}_a(q) &\equiv& \sum_{l=1}^{N}e^{ilq}S_l^{a},\quad a=1,2,3,
\end{eqnarray}
the ground state $|0\rangle$ has momentum $p_{s}=0,\pi$ and total spin
$S_{T}^{3}=S_{T}=NM(B)$, where $M(B)$ is the magnetisation.

The LSM construction \cite{LSM61,AL86} leads to gap-less excited states $|k\rangle$:
\begin{equation}\label{eq:2}
|k\rangle = {\bf U}^k |0\rangle,
 \quad
   {\bf U} \equiv\exp \left(-i\frac{2\pi}{N} \sum_{l=1}^{N} l S^{3}_{l}\right),
\end{equation}
with momenta
\begin{equation}\label{eq:3}
  p_k = p_{s} + k q_3(M), \quad q_3(M)\!\equiv\!\pi (1-2M),
\end{equation}
e.g. for $M=1/4$ one finds a four fold degeneracy of the ground state with
momenta $p_k = p_{s} + k \pi/2,\; k=0,1,2,3$. The magnetisation curve for the
system~(\ref{eq:1}) has no plateaus. The latter appear, if translational
invariance is broken, by adding to (\ref{eq:1}) a periodic perturbation with wave
vector $q$, e.g.:
\begin{eqnarray}\label{eq:5}
  \bar{\bf D}_{j}(q) &\equiv& \frac{1}{2}\left[{\bf D}_{j}(q) + {\bf D}_{j}(-q)
  \right],\\
\label{eq:14}
  {\bf D}_{j}(q)       &\equiv & 2\sum_{l=1}^{N} e^{iql} {\bf S}_{l} \cdot {\bf S}_{l+j},
  \quad j=1,2,\ldots.
\end{eqnarray}
So far the case $j=1$ has been studied in detail \cite{FGK+99}. A pronounced
plateau has been found in the magnetisation curve if the period
($q/\pi=1/2,1/6,1/3$) coincides with the \textit{first} soft mode ($k=1$),
$q=q_{3}(M)=\pi(1-2M)$, i.e. a perturbation $\bar{\bf D}_{1}(\pi/2)$ generates a
plateau at $M=1/4$. However, adding to (\ref{eq:1}) a perturbation $\bar{\bf
  D}_{1}(\pi)$ no plateau has been observed at $M=1/4$. In the latter case the
\textit{second} soft mode at $q=\pi$ ($k=2$) coincides with the period of the
perturbation.  Therefore, the question arises, why certain possibilities for the
formation of plateaus, which are allowed according to the quantisation rule of
Oshikawa \textit{et al.} \cite{OYA97}, are nevertheless not realized with a
given perturbation. One possible answer to this question has been given in
Ref.~\cite{FGK+99}: The efficiency of the mechanism to generate a plateau, by
means of a periodic perturbation ${\bf D}_{1}(q)$, crucially depends on the
magnitude of transition matrix elements $\langle n |{\bf D}_{1}(q) | 0 \rangle$ from the
ground state $|0\rangle$ to the low-lying excited states $| n \rangle$. For example for
$M=1/4$ the transition matrix elements $\langle n | {\bf D}_{1}(\pi/2) | 0 \rangle$ turn
out to be large whereas $\langle n | {\bf D}_{1}(\pi) | 0 \rangle$ are small. The effect
can be seen directly in the static dimer-dimer structure factor $\langle 0 |{\bf
  D}_{1}(q) {\bf D}_{1}(-q) | 0 \rangle$ for $M=1/4$ (cf. Fig.~4 in
Ref.~\cite{FGK+99}), which has a pronounced peak at $q=\pi/2$ but no peak for
$q=\pi$. Of course all these statements only hold for the nearest neighbour
Hamiltonian (\ref{eq:1}).

Indeed, Totsuka \cite{Tots98} has recently observed a magnetisation plateau at
$M=1/4$, which was created by adding to the Hamiltonian (\ref{eq:1}) a
perturbation $\bar{\bf D}_{1}(\pi)$ and a strong next-to-nearest neighbour
coupling.

In this paper we show that the magnetisation plateau at $M=1/4$ can be realized
as well with perturbations of period $q=\pi$, if the perturbation operator is
chosen properly. We will discuss in detail the situation with the operator
$\bar{\bf D}_{2}(q)$ defined in (\ref{eq:5}). Note, that both operators
$\bar{\bf D}_{1}(q)$ and $\bar{\bf D}_{2}(q)$ have the same momentum- and
spin-symmetry properties; they change the momentum by $\pm q$ and do not change
the total spin. They only differ in the isotropic spin-spin couplings, which
extend over nearest neighbours in $\bar{\bf D}_{1}(q)$ and over
next-nearest-neighbours in $\bar{\bf D}_{2}(q)$.

In Sec.~\ref{sec:signals-soft-mode} we compare the static structure factors,
$S_{j} \sim \langle 0 | {\bf D}_{j}(q) {\bf D}_{j}(-q) | 0 \rangle,\; j=1,2$, in the
presence of a magnetic field with magnetisation $M=1/4$.  In
Sec.~\ref{sec:magn-plat-induc} we study the magnetisation plateaus induced by
the periodic perturbation $\bar{\bf D}_{2}(q)$ at $q=\pi$.

The zig zag ladder with two (and three) legs -- recently investigated in
Ref.~\cite{CHP99} -- can be mapped on a 1D system with translation invariant
coupling over one and two (three) lattice spacings and a translation invariance
breaking coupling of the type $\bar{\bf D}_{2}(q)$. In
Sec.~\ref{sec:magn-plat-zig} we analyse the sequence of magnetisation plateaus,
which appears in the zig-zag ladder system.

It should finally be added that the operators and Hamiltonians refer to periodic 
boundary-conditions in leg directions. The DMRG results given in
Sec.~\ref{sec:magn-plat-zig}, however, have been obtained using open
boundary-conditions along the legs. 
%%%%%%%%%%%%%%%%%%%%%%%%%%%%%%%%%%%%%%%%%%%%%%%%%
%
\section{Signals of soft modes in static structure factors}
\label{sec:signals-soft-mode}
%
%%%%%%%%%%%%%%%%%%%%%%%%%%%%%%%%%%%%%%%%%%%%%%%%%
In this section we discuss some properties of the static structure factors $\langle 0
| {\bf D}_{j}(q) {\bf D}_{j}(-q) | 0 \rangle,\, j=1,2$, of the Hamiltonian
\eqref{eq:1}. In order to compute static structure factors we use 
periodic boundary conditions and exact Lanczos diagonalizations up to $N=24$
sites. 

Soft modes, as predicted by the LSM construction, can be seen
directly as zeros in the dispersion curve \cite{FGK+99}:
\begin{equation}
  \label{eq:6}
  \omega(q,M) = E(p_{s}+q,S+1)-E(p_{s},S),\quad M=S_{T}^{z}/N,
\end{equation}
where $E(p,S)$ are the lowest energy eigenvalues with total spin $S=S_{T}$ and
momentum $p$. The ground-state momentum $p_{s}$ in the sector with total spin
$S$ is known to be 0 or $\pi$\cite{YY66a}.  The zeros of \eqref{eq:6} -- in the
limit $N\to\infty $ -- appear at the soft mode momenta $q=q^{(k)}(M)$. For example,
at $M=1/4$ three zeros at $q/\pi=0,1/2,1$ emerge in the dispersion curve for the
Hamiltonian (\ref{eq:1}) (cf. Fig.~3 of Ref.~\cite{FGK+99}).

\begin{figure}[ht]
\centerline{
\epsfig{file=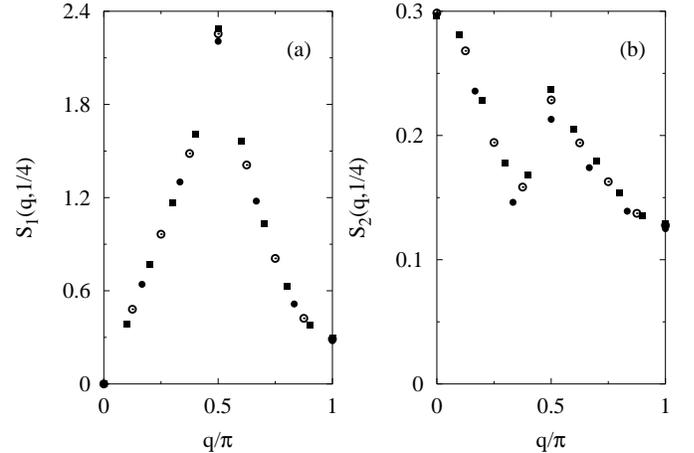,width=6.2cm,angle=-90}}
\caption{Comparison of the $q$-dependence of the static structure factors
  \eqref{eq:9} at $M=1/4$ for the dimer operators (\ref{eq:5}) over nearest
  [ (a) $j=1$] and next-nearest-neighbour [ (b) $j=2$] couplings in a model with
  Hamiltonian (\ref{eq:1}). The different symbols belong to system sizes: $N=20\,
  (\blacksquare),\,16 (\circ),\, 12\, (\bullet)$.}
\label{fig:h1-sdd-m1d4}
\end{figure}

The operators $\mathbf{D}_{j}(q),\, j=1,2$, defined in Eq.~(\ref{eq:14}),
commute with the total spin squared $\mathbf{S}_{T}^{2}$ and the dispersion
curve (\ref{eq:6}) describes the lowest-lying excitations, which can be reached
with these operators. The transition matrix element
\begin{equation}
  \label{eq:8}
  \langle n | \mathbf{D}_{j}(q) | 0 \rangle, \quad j=1,2,
\end{equation}
from the ground state $|0\rangle$ with total spin $S=S_{T}$ and momentum $p_{s}$
to the excited states $|n\rangle$ with momentum $p=p_{s}+q$ enter in the
corresponding static structure factor:
\begin{equation}
  \label{eq:9}
  S_{j}(q,M) \equiv \frac{1}{N} \sum_{n} 
  (1-\delta_{n0})|\langle n | \mathbf{D}_{j}(q) | 0 \rangle|^{2}
\end{equation}
The signals of the soft modes in the static structure factor therefore measure
the magnitude of the transition matrix elements (\ref{eq:8}). In
Figs.~\ref{fig:h1-sdd-m1d4}(a) and \ref{fig:h1-sdd-m1d4}(b) we compare the
$q$-dependence of the static structure factors $S_{j}(q,M\!=\!1/4)$.
%%%%%%%%%%%%%%%%%%%%%%%%%%%BEGIN-FIGURE%%%%%%%%%%%%%
%%%%%%%%%%%%%%%%%%%%%%%%%%%END-FIGURE%%%%%%%%%%%%%%
Indeed we observe remarkable differences. We find a pronounced peak in both
structure factors $S_{j}(q,1/4),\, j\!=\!1,2$ at the soft mode
$q^{(1)}(1/4)\!=\!\pi/2$. The size of the peaks, however, differ by an order of
magnitude. There is no peak at the second soft mode $q^{(2)}(1/4)=\pi$. Note the
different behaviour of the structure factor. For $q\to 0$: $S_{1}(q,1/4)$
converges to zero whereas $S_{2}(q,1/4)$ approaches a maximum $\lesssim 0.3$.  This
feature will play a crucial role, if we add to (\ref{eq:1}) a periodic
perturbation $\delta \cdot \bar \mathbf{D}_{j}(\pi),\;j=1,2$ of strength $\delta$.  The
perturbation is invariant under translations by two lattice spacings.
Eigenstates of the perturbed Hamiltonian are constructed by a superposition of
momentum eigenstates with $p=0$ and $p=\pi$. The reduction of the Brioullin zone
is taken into account in a modification of the dimer operators:
\begin{equation}\label{eq:23}
 \mathbf{D}_{j}^{\kappa}(q) \equiv 
      2 \sum_{l=0}^{N/2-1}e^{iq2l}\ {\bf S}_{2l+\kappa} \cdot {\bf S}_{2l+j+\kappa},
      \quad 
      \begin{cases}
        j=1,2 \\ \kappa=0,1.
      \end{cases}
\end{equation}
The corresponding structure factors, defined by:
\begin{eqnarray}
  \label{eq:24}
  S_{j}^{\kappa}(q,M) &\equiv& \frac{1}{N}\left[ 
  \langle 0 |  \mathbf{D}_{j}^{\kappa}(q)  \mathbf{D}_{j}^{\kappa}(-q) | 0 \rangle - 
 \left|\langle 0 |  \mathbf{D}_{j}^{\kappa}(q)  | 0 \rangle\right|^{2} \right],\nonumber \\
\end{eqnarray} 
are symmetric under the mapping $q\to\pi-q$. 

\begin{figure}[ht]
  \centerline{
\epsfig{file=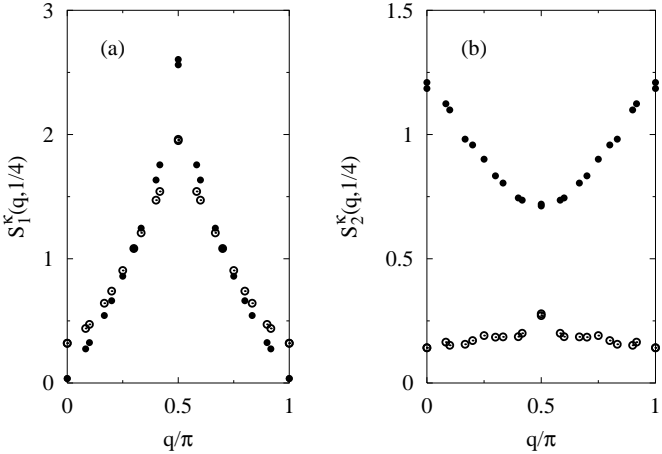,width=6.2cm,angle=-90}}
\caption{Comparison of the $q$-dependence of the static structure factor
  \eqref{eq:24} at $M\!=\!1/4$ for the dimer operators \eqref{eq:23} in a model
  with Hamiltonian \eqref{eq:1} and perturbation $\delta \cdot
  \mathbf{D}_{j}^{\kappa}(\pi),\; \kappa\!=\!0 (\bullet), 1(\circ);\, j\!=\!1$(a)$,
  2$(b). The perturbation strength is $ \delta\!=\!0.5$, the system sizes are
  $N=20,24$.}
\label{fig:sddpert}
\end{figure}

In Fig.~\ref{fig:sddpert}(a) we present the static structure factor
$S_{1}^{\kappa}(q,1/4),\, \kappa\!=\!0,1$ -- obtained from the ground state of the
Hamiltonian (\ref{eq:1}) with perturbation $\delta \cdot \bar \mathbf{D}_{1}(\pi),\;
\delta=0.5$.
%%%%%%%%%%%%%%%%%%%%%%%%%%%BEGIN-FIGURE%%%%%%%%%%%%%
%%%%%%%%%%%%%%%%%%%%%%%%%%%END-FIGURE%%%%%%%%%%%%%%
In both structure factors we only find a peak at $q=\pi/2$ but no peak at
$q=0,\pi$.  The situation for the static structure factors $S_{2}^{\kappa}(q,1/4),\,
\kappa\!=\!0,1$ is shown in Fig.~\ref{fig:sddpert}(b). Here the ground state has
been computed for Hamiltonian (\ref{eq:1}) with a perturbation $\delta\cdot \bar
\mathbf{D}_{2}(\pi),\; \delta=0.5$.  Note the different behaviour of the two
structure factors $S_{2}^{\kappa}(q,1/4),\; \kappa=0,1$.  $S_{2}^{1}(q,1/4)$ has its
maximum at $q=0,\pi$, whereas $S_{2}^{2}(q,1/4)$ has a maximum at $q=\pi/2$. We
therefore expect that the operator $\bar \mathbf{D}_{2}(\pi)$ generates a plateau
at $M=1/4$, whereas the operator $\bar \mathbf{D}_{1}(\pi)$ does not.

%%%%%%%%%%%%%%%%%%%%%%%%%%%%%%%%%%%%%%%%%%%%%%%%%
%
\section{Magnetisation plateaus induced  by periodic perturbations}
\label{sec:magn-plat-induc}
%
%%%%%%%%%%%%%%%%%%%%%%%%%%%%%%%%%%%%%%%%%%%%%%%%%
Periodic perturbations of the type $\bar\mathbf{D}_{j}(q)$ -- added to the
nearest neighbour Hamiltonian (\ref{eq:1}) -- generate a characteristic sequence
of plateaus in the magnetisation curve. We dicsuss the same Hamiltonian for
which we have computed the static structure factor in Sec.~2.2. For this reason
we also apply periodic boundary conditions in this section.

\begin{figure}[ht]
  \centerline{
\epsfig{file=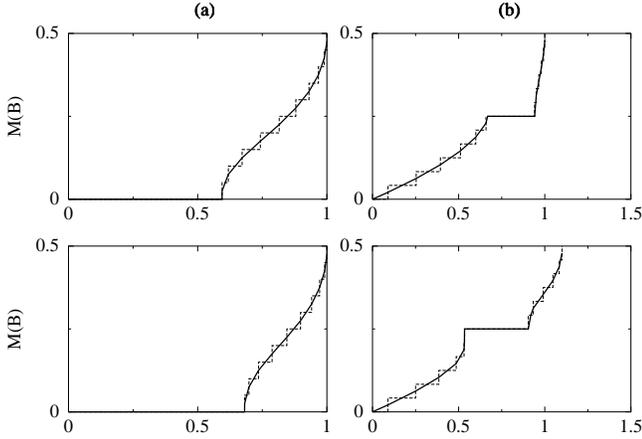,width=64mm,angle=-90}}
\caption{Comparison of the plateau evolution in the magnetisation curves of a
  model with Hamiltonian \eqref{eq:1} and perturbation $\delta \cdot \bar
  \mathbf{D}_{j}(\pi), j=1 (a), 2 (b),\, \delta=0.5,0.6$ (from top to
  bottom).  The solid lines represent the midpoint magnetisation curves,
  together with extrapolated values of the upper and lower critical field
  $B^{U}$ and $B^{L}$, respectively, deduced from system sizes
  $N=8,12,\ldots,24$.}
\label{fig:maghd1hd2}
\end{figure}

The possible position of the plateaus is given by the quantisation rule of
Oshikawa \textit{et al.}  \cite{OYA97}.  However, whether or not a plateau
really appears, crucially depends on the type of the perturbation operator. As
an example we compare in Fig.~\ref{fig:maghd1hd2} the evolution of the
magnetisation plateaus generated by the operators $\bar\mathbf{D}_{1}(\pi)$ and
$\bar\mathbf{D}_{2}(\pi)$.

The magnetisation curves -- generated with $\bar\mathbf{D}_{1}(\pi)$, cf.
Fig.~\ref{fig:maghd1hd2} column (a) -- show a plateau at $M=0$, rapidly
increasing with the strength $\delta$ of the perturbation. This is the well known
gap induced by dimerisation. There is no plateau at $M=1/4$ in the whole $\delta$
range ($0 <\delta \lesssim 0.7$).

\begin{figure}[ht]
  \centerline{ 
\epsfig{file=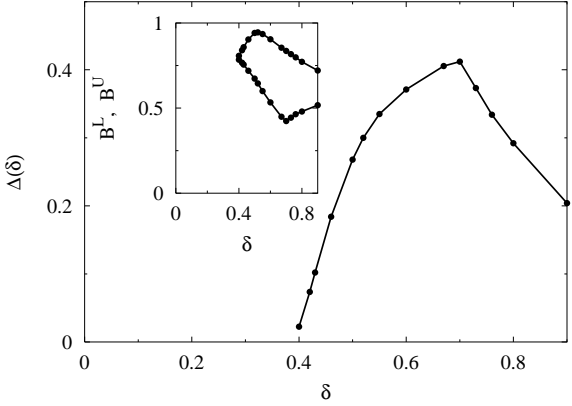,width=64mm,angle=-90}}
\caption{The $\delta$-evolution of the $M=1/4$ plateau width
  $\Delta(\delta)=B^{U}-B^{L}$ in a model with Hamiltonian \eqref{eq:1} and a
  perturbation $\delta \cdot \bar \mathbf{D}_{2}(\pi)$, deduced from system sizes
  $N=12,16,20,24$.}
\label{fig:plateauhd2}
\end{figure}

In contrast the magnetisation curves -- generated with $\bar\mathbf{D}_{2}(\pi)$,
[cf.  Fig.~\ref{fig:maghd1hd2} column (b)] -- have no plateau at $M=0$. For $\delta\gtrsim0.4$ a
plateau appears at $M=1/4$. A finite-size analysis of the plateau width yields
the $\delta$-evolution shown in Fig.~\ref{fig:plateauhd2}.

Note in particular, the drastic change in the plateau width at $\delta=0.7$. This
feature is associated with a change in the ground-state quantum numbers in the
fixed $S_{T}^{z}$-sectors. For $\delta<1/2$ all ground states ($0\leq M \leq 1/2$) have
the standard momentum $0,\pi$. The first change happens in the sector
$M\!=\!1/2\!-\!1/N$ at $\delta\!=\!1/2$, where the ground state is degenerate with
momentum $p\!=\!0,\pi$ and $p\!=\!\pm \pi/2$.

For larger values of $\delta$ $(\delta>1/2)$ we observe a different ground-state
behaviour in the sectors with:
\begin{equation}
  \label{eq:25}
  0\leq M \leq M_{0}(\delta), 
\quad \mbox{and} \quad
  M_{0}(\delta)\leq M \leq 1/2.
\end{equation}
In the first regime of (\ref{eq:25}) the ground state has still momentum $p=0,\pi$.
In contrast, in the second regime of (\ref{eq:25}) the ground-state momentum
alternates between $p=0,\pi$ and $p=\pm \pi/2$. The magnetisation $M_{0}(\delta)$, which
separates the two regimes passes the plateau $M\!=\!1/4$ exactly at $\delta=0.7$.
%%%%%%%%%%%%%%%%%%%%%%%%%%%%BEGIN-FIGURE%%%%%
%%%%%%%%%%%%%%%%%%%%%%END-FIGURE%%%%%%%%%%%%

%%%%%%%%%%%%%%%%%%%%%%%%%%%%BEGIN-FIGURE%%%%%
%%%%%%%%%%%%%%%%%%%%%END-FIGURE%%%%%%%%%%%%

The plateau structures in the magnetization curves of chains with periodic
perturbations can be seen already on rather small systems ($N\approx 24$). 
%%%%%%%%%%%%%%%%%%%%%%%%%%%%%%%%%%%%%%%%%%%%%%%%%
%
\section{Magnetisation plateaus in zig-zag ladders}
\label{sec:magn-plat-zig}
%
%%%%%%%%%%%%%%%%%%%%%%%%%%%%%%%%%%%%%%%%%%%%%%%%%
The magnetic properties of zig-zag ladders have been recently investigated by
Cabra, Honecker and Pujol \cite{CHP99} by means of bosonization techniques and
numerical analysis. In order to understand the appearance of plateaus in the
magnetisation curve, a mapping of the ladder system onto a 1D spin chain with
couplings over short range distances is useful. For \textit{normal} spin ladders
with $l$ legs this mapping leads to couplings over nearest neighbour and to
couplings over $l$ lattice sites \cite{FKM99}. The couplings over $l$ lattice
sites appear along the legs, whereas nearest neighbour couplings form the rungs
of the ladder. At the endpoints of the rungs nearest neighbour bonds have to
vanish to avoid the appearance of diagonal couplings in the ladder. A Fourier
analysis of the translation invariance breaking terms:
\begin{equation}
  \label{eq:10}
  \sum_{q} \delta_{q} \bar D_{1}(q),
\end{equation}
leads to a prediction of magnetisation plateaus at:
\begin{equation}
  \label{eq:11}
  M = M_{l}^{Z} \equiv \frac{1}{2} - \frac{Z}{l},
\end{equation}
where $Z$ is integer and runs over the sequence
\begin{equation}
  \label{eq:12}
  Z =  \begin{cases}
    1,2,3,\ldots,l/2     &: l \mbox{ even} \\
    1,2,3,\ldots,(l-1)/2 &: l \mbox{ odd}.
  \end{cases}
\end{equation}
The prediction is based on the assumption that a plateau only appears if one of
the wave vectors $q$ in the Fourier analysis (\ref{eq:10}) of the perturbation
coincides with the \textit{first} soft mode $(k=1)$.

In the following subsections we consider different realizations of zig-zag spin
ladders, which have been discussed recently by several authors
\cite{CHP99,WSME99,Kole99}. We map those systems to 1D systems with appropriate
couplings over $j=1,2,3$ neighbours. For some of the following systems we expect
more than one plateau in the magnetization curve, therfore we use in this
section DMRG calculations, which also means we apply open boundary conditions, to
get a finer resolution, i.e. more steps in the finite-size magnetization curves.
%%%%%%%%%%%%%%%%%%%%%%%%%%%%%%%%%%%%%%%%%%%%%%%%%
\subsection{The two-leg zig-zag ladder}
\label{sec:two-leg-zig}
\begin{figure}[ht]
\centerline{
\epsfig{file=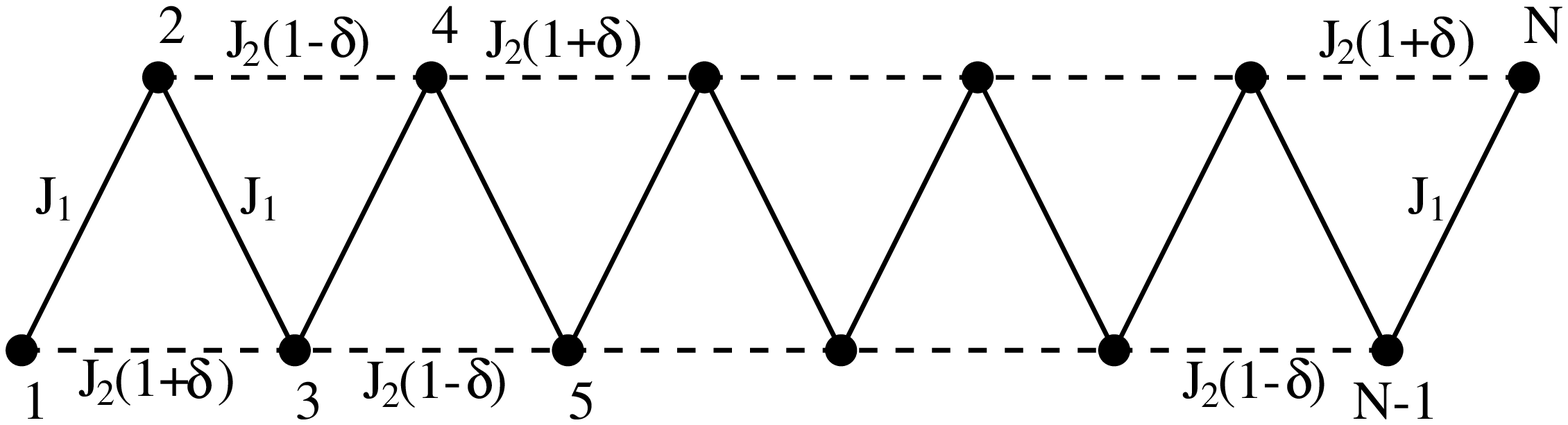,width=8.4cm}}
\caption{Mapping of the two-leg zig-zag ladder onto a 1D system with Hamiltonian 
  \eqref{eq:15}. The solid lines represent the nearest neighbour interactions
  with coupling $J_{1}$ and the dashed lines the next-nearest-neighbour
  interactions with couplings $J_{2}(1\pm \delta)$.}
\label{fig:twoleg}
\end{figure}

\begin{figure}[ht]
\centerline{
\epsfig{file=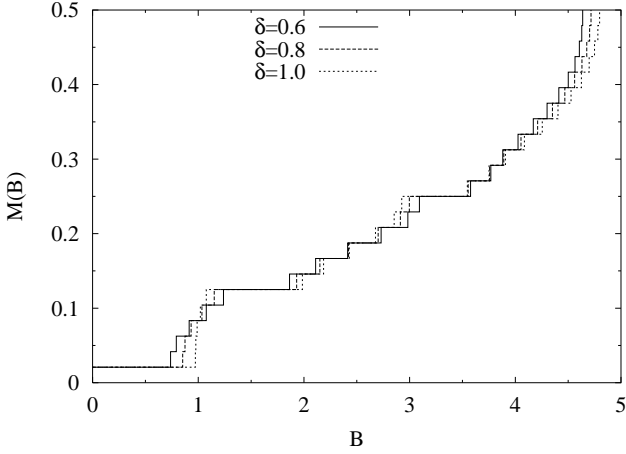,width=6.2cm,angle=-90}}
\caption{Magnetisation curve of the two leg zig-zag ladder with 
  Hamiltonian~\eqref{eq:13} with couplings:
  $J_{1}=1,\;J_{2}=2,\;\delta=0.6,0.8,1.0$. The system size is $N=48$.}
\label{fig:twoleg-zz}
\end{figure}

%%%%%%%%%%%%%%%%%%%%%%%%%%%%%%%%%%%%%%%%%%%%%%%%%
The mapping of the two leg zig-zag ladder onto a 1D system with short range
couplings is shown in Fig.~\ref{fig:twoleg}.
%%%%%%%%%%%%%%%%%%%%%%%%%%%%BEGIN-FIGURE%%%%%
%%%%%%%%%%%%%%%%%%%%%%%END-FIGURE%%%%%%%%%%%%
The Hamiltonian can be written as:
\begin{eqnarray}
  \label{eq:15}
    {\bf H} &=& J_1{\bf H}_{1}+ J_2(1+\delta){\bf H}_{+\delta}+
    J_2(1-\delta){\bf H}_{-\delta}, 
\\
  {\bf H}_{\pm \delta} &\equiv& 2\sum_{l=1}^{N}
  \left[{\bf S}_{4l} \cdot {\bf S}_{4l\pm 2} + {\bf S}_{4l+1} \cdot {\bf S}_{4l\pm
      2+1}\right] .
\end{eqnarray}
The Fourier analysis \eqref{eq:10} of the translation invariance breaking terms
yields in this case:
\begin{eqnarray}
  \label{eq:13}
  {\bf H} &=& \sum_{j=1,2}J_j{\bf H}_{j}+
  \delta J_{2}\sqrt{8} \sum_{l=1}^{N}
  \cos\left(\frac{\pi l}{2}-\frac{\pi}{4}\right) 
  {\bf S}_{l} \cdot {\bf S}_{l+2}.\nonumber \\
\end{eqnarray}
Therefore we expect magnetisation plateaus, if the wave vector $q/\pi=1/2,3/2$
meets the first and second soft mode:
\begin{eqnarray}
  \label{eq:16}
  q =\frac{\pi}{2} \quad &:& \quad  
  M_{4k}^{1} = \frac{1}{2} - \frac{1}{4k}, \quad \mbox{ for } k=1,2, \\
 q =\frac{3\pi}{2} \quad &:& \quad  
  M_{4k}^{3} = \frac{1}{2} - \frac{3}{4k}, \quad \mbox{ for } k=2.
\end{eqnarray}
We have looked in particular for plateaus at $M_{8}^{3}=1/8$ and $M_{8}^{1}=3/8$
induced by the second soft mode $k=2$. The situation at $J_{1}=1,\; J_2=2$
and $\delta=0.6, 0.8, 1.0$ is shown in Fig.~\ref{fig:twoleg-zz}, where the emergence
of a plateau at $M=1/8$ is visible. The effect disappears if we change the ratio
$\alpha=J_{1}/J_{2}$ in both directions $\alpha<1/2$ and $\alpha> 1/2$.
%%%%%%%%%%%%%%%%%%%%%%%%%%%%BEGIN-FIGURE%%%%%
%%%%%%%%%%%%%%%%%%%%%%%%%%%%%END-FIGURE%%%%%%%%%%%%
%%%%%%%%%%%%%%%%%%%%%%%%%%%%%%%%%%%%%%%%%%%%%%%%%
\subsection{Three-leg zig-zag ladder with periodic rung couplings}
\label{sec:Three-leg-zig}
\begin{figure}[ht]
\centerline{
\epsfig{file=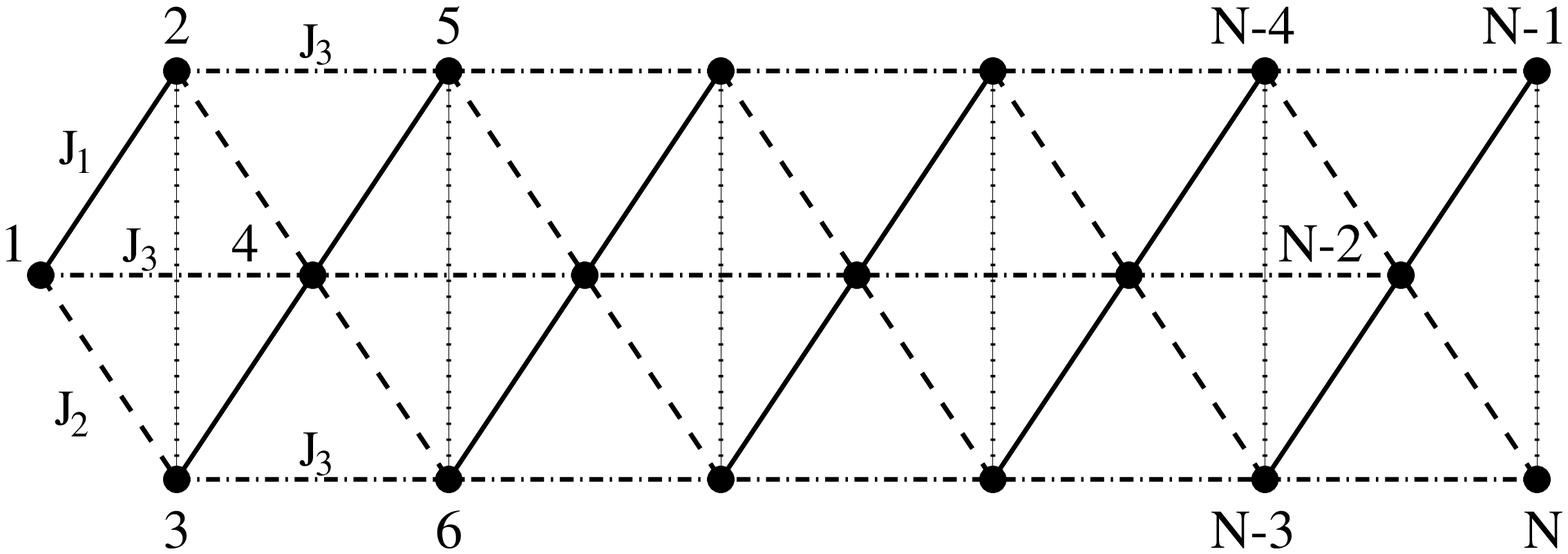,width=8.8cm}}
\caption{Mapping of the three leg zig-zag ladder with periodic rung couplings
  onto a 1D system with Hamiltonian \eqref{eq:17}. The solid lines represent
  nearest neighbour interactions with couplings $J_{1}$, the dashed lines
  next-nearest-neighbour interactions with couplings $J_{2}$, and the
  dashed-dotted lines interactions over third neighbour interactions with
  couplings $J_{3}$. The dotted lines denote interactions with nearest neighbour 
  couplings $J_{1}$, as they are enforced by periodic boundary conditions along
  the rungs.}
\label{fig:threeleg}
\end{figure}

%%%%%%%%%%%%%%%%%%%%%%%%%%%%%%%%%%%%%%%%%%%%%%%%%
The mapping of the three-leg zig-zag ladder onto a 1D system with short range
couplings is shown in Fig.~\ref{fig:threeleg}.
%%%%%%%%%%%%%%%%%%%%%%%%%%%%BEGIN-FIGURE%%%%%
%%%%%%%%%%%%%%%%%%%%%%%%%%%%%END-FIGURE%%%%%%%%%%%%
The corresponding Hamiltonian can be rewritten as:
\begin{eqnarray}
  \label{eq:17}
    {\bf H}  &=& \sum_{j=1,3}J_{j}{\bf H}_{j} + 
    \frac{2}{3}J_{2}\left[
      {\bf H}_{2} - 
     2 \sum_{l=1}^{N} \cos\left(\frac{2\pi}{3}l\right) {\bf S}_{l} \cdot {\bf S}_{l+2}
    \right]. \nonumber \\
\end{eqnarray}
The last term on the right-hand side breaks the translation invariance of the 1D
system and we therefore expect magnetisation plateaus for
\begin{equation}
  \label{eq:19}
    M_{3}^{1} = \frac{1}{6}, \quad 
    M_{6}^{1} = \frac{1}{3}, \quad 
    M_{9}^{2} = \frac{7}{18},
\end{equation}
if the wave vector $q=2\pi/3$ meets the first ($k=1$), second ($k=2$) and third
($k=3$) soft mode, respectively. 

\begin{figure}[ht]
  \centerline{
\epsfig{file=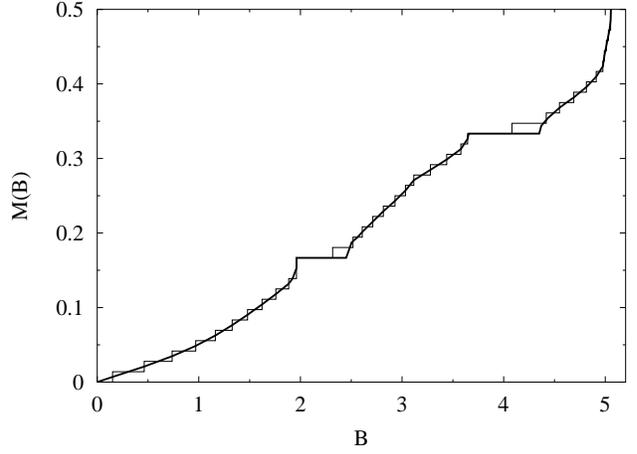,width=6.2cm,angle=-90}}
\caption{Magnetisation curve of the three leg zig-zag ladder with periodic rung
  couplings and Hamiltonian \eqref{eq:17} with couplings:
  $J_{1}=1.5,\;J_{2}=1.5$. The system size is $N=72$. The solid line is the
  midpoint magnetisation curve. The behaviour at the end of the plateaus is
  extrapolated from data of systems $N=24,36,\ldots,96$. The finite-size
  effects are largest in the region around the upper critical field of the plateaus.}
\label{fig:mag-three-or}
\end{figure}

We have computed the magnetisation curves with periodic rung couplings and open
boundary-conditions along the legs for the following values:
\begin{equation}
  \label{eq:20}
  J_{1}=\frac{3}{2}J_{3}, \quad J_{2}=\frac{3}{2}J_{3},
\end{equation}
and system sizes: $N=3\times L, \; L=8,12,16,20,24$.  It turns out, that the midpoint
extrapolation {\`a} la Bonner and Fisher \cite{BF64} still shows a surprisingly large
finite-size effects for the open boundary conditions. However, as can be seen
from Fig.~\ref{fig:mag-three-or}, there is still a clean signal for two plateaus
at $M=1/6$ and $M=1/3$.

In Ref.~\cite{CHP99}, the magnetisation curve has been computed for the same set
of couplings \eqref{eq:20}, but with different boundary-conditions along the
rungs, which they call periodic boundary-conditions of type A, B or C. No
plateau at all is found for type B, one plateau at $M=1/6$ is found for type A
and C.  This is a first indication that the formation of plateaus critically
depends on the boundary-conditions along the rungs.

%%%%%%%%%%%%%%%%%%%%%%%%%%%%BEGIN-FIGURE%%%%%
%%%%%%%%%%%%%%%%%%%%%%%%%%%%%END-FIGURE%%%%%%

%%%%%%%%%%%%%%%%%%%%%%%%%%%%%%%%%%%%%%%%%%%
\subsection{Three-leg ladder with open rung couplings}
\label{sec:Three-leg-ladder}
%%%%%%%%%%%%%%%%%%%%%%%%%%%%%%%%%%%%%%%%%%%%%%%%%
The mapping of this system onto a 1D system with short range couplings can be
seen again from Fig.~\ref{fig:threeleg}. We only have to remove the dotted
nearest neighbour bonds, which implement the periodic rung couplings. This
changes the Hamiltonian in the following manner:
\begin{eqnarray}
  \label{eq:21}
  {\bf H}  &=&  \frac{2}{3}
  \sum_{j=1,2}
   J_{j}\left[
     {\bf H}_{j} - 
      2\sum_{l=1}^{N} \cos\left[\frac{2\pi}{3}(l\!+\!2\!-\!j)\right] {\bf S}_{l} \cdot {\bf S}_{l+j} 
    \right] +
     \nonumber \\ && + J_{3}{\bf H}_{3} .
\end{eqnarray}
The two Hamiltonians (\ref{eq:17}) and (\ref{eq:21}) differ in the relative
weight of the translation symmetry breaking terms. 

\begin{figure}[ht]
  \centerline{
\epsfig{file=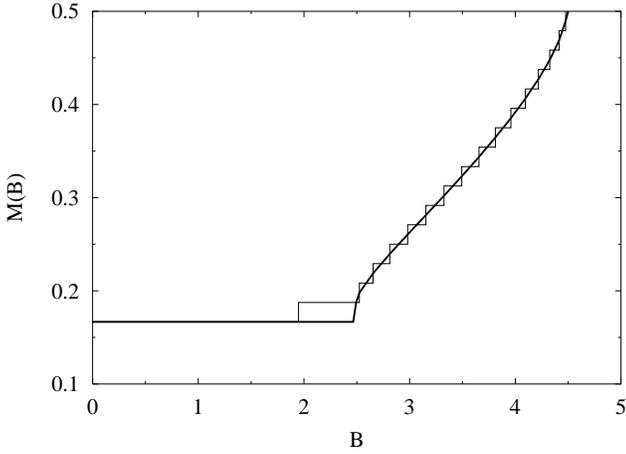,width=6.2cm,angle=-90}}
\caption{Magnetisation curve of the three leg zig-zag ladder with open rung
  and leg couplings and Hamiltonian~\eqref{eq:21} with couplings:
  $J_{1}=J_{2}=3/2,\;J_{3}=-1$. The solid line represents the midpoint
  magnetisation, extrapolated from system sizes $N=24-72$. The thin line is the
  magnetisation curve for $N=48$. }
\label{fig:mag-three-os}
\end{figure}

It was found already in Ref.~\cite{CHP99} that for the couplings (\ref{eq:20})
$(J_{1}=3/2,\; J_{2}=3/2,\; J_{3}=1)$ the magnetisation curve for the
Hamiltonian (\ref{eq:21}) evolves two pronounced plateaus at $M=1/6$ and at
$M=1/3$ generated by the first and second soft mode, respectively.  This is
confirmed in our computation. Switching from antiferromagnetic to ferromagnetic
leg coupling $(J_{3}=-1)$ we find spontaneous magnetisation at $M=1/6$ whereas
the plateau at $M=1/3$ disappears (Fig.~\ref{fig:mag-three-os}).
%%%%%%%%%%%%%%%%%%%%%%%%%%%%BEGIN-FIGURE%%%%%
%%%%%%%%%%%%%%%%%%%%%%%%%%%%%END-FIGURE%%%%%%

%%%%%%%%%%%%%%%%%%%%%%%%%%%%%%%%%%%%%%%%%%%%%%%%%
\subsection{The Kagom{\'e} like \textit{three spin ladders }}
\label{sec:Kagome-like-three}
%%%%%%%%%%%%%%%%%%%%%%%%%%%%%%%%%%%%%%%%%%%%%%%%%
This system has been studied recently in Ref.~\cite{WSME99} for reasons which we
explain below. Its couplings are defined by removing from
Fig.~\ref{fig:threeleg} the middle leg and the dotted vertical lines. The
mapping onto the 1D system with short range couplings then leads to the
Hamiltonian:
\begin{eqnarray}
  \label{eq:27}
     {\bf H}  &=& 
     \frac{2}{3}\sum_{r=1}^{3} J_{r}{\bf H}_{r}
     - \nonumber \\  
     &&\hspace{2mm}
     \frac{4}{3}\left(
       J_{1}\sum_{l=1}^{N}\cos\left[\frac{2\pi}{3}(l+1)\right]
       {\bf S}_{l} \cdot {\bf S}_{l+1} +     
     \right.  \nonumber  \\
      && \hspace{2mm}
       J_{2} \sum_{l=1}^{N}\cos\left[\frac{2\pi}{3}l\right]
       {\bf S}_{l} \cdot {\bf S}_{l+2} +  
      \nonumber  \\
      && \hspace{2mm}
       \left. J_{3}\sum_{l=1}^{N}\cos\left[\frac{2\pi}{3}(l-1)\right]
         {\bf S}_{l} \cdot {\bf S}_{l+3},
       \right).
\end{eqnarray}
Equation (\ref{eq:27}) differs from Eq.~(\ref{eq:21}) in the additional
translation symmetry breaking term over three lattice spacings. The Kagom{\'e} like
\textit{three leg ladder} is interesting because it shares with the two
dimensional Kagom{\'e} lattice the property that there is a high density of
low-lying singlets. A different Kagom{\'e} like three leg ladder system has been
analysed recently by Azaria \textit{et al.} \cite{ALN98}.

The Kagom{\'e} lattice itself seems to have a singlet-triplet gap. The authors of
Ref.~\cite{WSME99} tried to find out whether such a gap (i.e. a plateau at $M=0$)
exists as well for the Kagom{\'e} like \textit{three spin ladder}.  Extrapolation
from DMRG results for systems up to 120 sites show that the system is gap-less
for values of the leg spin coupling $J$ in the interval $0.5 < 2J_{3} < 1.25, \ 
J_{1}=1/2$. Note, that the translation invariance breaking terms in
(\ref{eq:27}) do not generate a plateau at $M=0$ but at $M=1/6$ and $M=1/3$.

\begin{figure}[ht]
  \epsfig{file=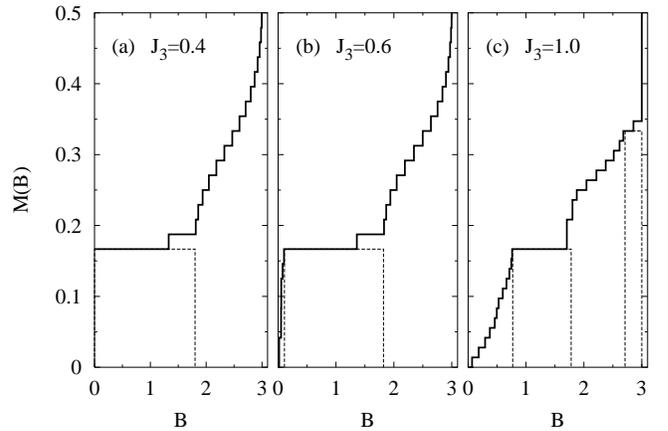,width=6.2cm,angle=-90}
\caption{Magnetisation curve of the Kagom{\'e} like \textit{three spin ladder}
 with Hamiltonian~\eqref{eq:27} and  couplings:
  $J_{1}=J_{2}=1,\;J_{3}=0.4,0.6,1.0$. The system size is $N=48$ for (a) and (b)
  and due to the larger finite-size effects $N=72$ for (c). The dashed lines
  show the extrapolted postions of the plateau ends deduced from Lanczos
  diagonalizations with periodic boundary conditions.}
\label{fig:kagome}
\end{figure}

Our results for fixed couplings $J_{1}=J_{2}=1.0$ and increasing couplings $J_{3}=0.4,
0.6, 1.0$ are shown in Fig.~\ref{fig:kagome}. For $J_{3}=0.4$ we find
spontaneous magnetisation at $M=1/6$. This phenomenon is a consequence of the
Lieb-Mattis theorem \cite{LM62}, as was pointed out by the authors of
Ref.~\cite{WSME99}. In condensed matter physics this phenomenon is called
\textit{ferrimagnetism}. The magnetisation curve for $J_{3}=0.6$ starts with a
steep increase at $B=0$ and reaches quickly the plateau at $M=1/6$ for a small
value of $B\approx0.1$. Going to larger values of the coupling $J_{3}=1.0$, the slope
of the magnetisation curve and the plateau width at $M=1/6$ are reduced. At the
same time, we expect the appearance of a small plateau at $M=1/3$. Beyond this
point the system jumps into the saturation magnetisation. This phenomenon is
called meta-magnetism~\cite{GMK98}. 
%%%%%%%%%%%%%%%%%%%%%%%%%%%%BEGIN-FIGURE%%%%%
%%%%%%%%%%%%%%%%%%%%%%%%%%%%%END-FIGURE%%%%%%

We finally want to give some comments on the strong finite-size behavior
of the presented DMRG calculations of magnetization curves especially
appearing at the upper  critical fields of the shown magnetization
plateaus. Reanalyzing the 3-leg zig-zag and Kagom{\'e} like ladders of this
section for system sizes $N=12,18,24$ and periodic leg boundary 
conditions applying standard Lanczos techniques led to a straight
affirmation of the presented extrapolations. It moreover showed that the
strong finite-size effects at the upper plateau edges are no genuine
additional features but a simple consequence of the open leg boundary
conditions used for the DMRG calculations (see e.g. plateau boundaries
additionally given in Fig. \ref{fig:kagome}). In addition, the evaluations
with periodic leg boundary conditions could make clear the existence
of an additional plateau ($m=1/3$) for the Kagom{\'e} like 3-leg ladder
shown in Fig. \ref{fig:kagome} ($c$), i.e. $J_3=1.0$. Again, at the
upper critical field strong finite-size corrections hinder the
identification of the $m=1/3$ plateau on the basis of the shown DMRG
results. For the two other cases --$J_3=0.4,\,0.6$-- additional
plateaus could be excluded. It remains to be summarized that 
evaluations with both types
of leg boundary conditions agree --if possible to obtain-- in the 
thermodynamic limits of the considered quantities. Periodic boundary
conditions, however, show much smaller and more controlled behavior
of finite-size effects while the open boundary conditions require
considerably larger system sizes for an equal quality of extrapolation.

%%%%%%%%%%%%%%%%%%%%%%%%%%%%%%%%%%%%%%%%%%%%%%%%%
%
\section{Discussion and Conclusions}
\label{sec:conclusions}
%
%%%%%%%%%%%%%%%%%%%%%%%%%%%%%%%%%%%%%%%%%%%%%%%%%
The Lieb-Schultz-Mattis construction of gap-less excited states (\ref{eq:2}) in
quasi one-dimensional spin-1/2 quantum spin systems demands translation
invariance and short range couplings. To our knowledge systems which satisfy
these conditions have no plateaus in their magnetisation curve for $M>0$. In
this paper we have studied the effect of a modulation of the couplings over
$j=1,2,3$ sites [cf. Eq.~\eqref{eq:5}] on the magnetisation process.

According to the quantisation rule of Oshikawa, Yamanaka and Affleck, we expect
plateaus if the wave number $q$ of the modulation coincides with one of the
momenta of the soft modes $q^{(k)}=k\pi(1-2M),\ k=1,2,\ldots$. We have found, that
the modulation of the nearest neighbour coupling [with the operator ${\bf \bar
  D}_{1}(q)$] generates one plateau at $M=1/2-q/\pi$, i.e. if $q$ coincides with
the first soft mode $q^{(1)}(M)$.

For $q=\pi$, the modulation of the nearest neighbour coupling with the operator
${\bf \bar D}_{1}(\pi)$ leads to the well-known singlet-triplet gap, i.e. a
plateau at $M=0$ -- which opens with the strength $\delta$ of the perturbation
as $\delta^{2/3}$\cite{CF79,FKM98}.  

The modulation of the next-nearest-neighbour coupling with ${\bf \bar
  D}_{2}(\pi)$ does not affect substantially the magnetisation process if $\delta<
0.4$. There is neither a plateau at $M=0$ nor at $M=1/4$. For $\delta>0.4$, however,
a plateau opens rapidly at $M=1/4$ and shrinks again for $\delta > 0.7$. We have
found that this effect is correlated with a change in the quantum numbers of the
ground state.

Spin ladders with $l$ legs can be mapped on one dimensional systems with
modulated short range couplings. Of course these mappings are not unique. There
are many possibilities to put a ring with nearest neighbour couplings on a
ladder in such a way, that each site is passed once. The links which do not lie
on the ring, define further reaching and modulated couplings. However, these
mappings become unique, if we postulate that the range of the couplings is
minimal. This means for normal ladders with $l$-legs, that the corresponding 1D
system only contains modulated nearest neighbour couplings $[{\bf \bar
  D}_{1}(q)]$ and (translation invariant) couplings over $l$ lattice sites. Here
magnetisation plateaus appear if the wave vectors $q$ of the magnetisation
plateaus meet the \textit{first} soft mode. The situation is different if we map
zig-zag ladders on quasi 1D systems. Modulations of the next-nearest-neighbour
couplings $[{\bf \bar D}_{2}(q)]$ emerge, which generate magnetisation plateaus
if the wave vector $q$ of the modulation meets either the first or the
\textit{second} soft mode.

Finally, we studied the magnetisation process of three leg zig-zag ladders with
various boundary-conditions along the rungs. The boundary-conditions change the
weight of the terms which modulate the couplings over one, two and three lattice 
spacings in the 1D Hamiltonian. This again affects the formation of plateaus at
$M=1/6$ and $M=1/3$, respectively.

%%%%%%%%%%%%%%%%%%%%%%%%%%%%%%%%%%%%%%%%%%%%%%%%%
%
%\begin{appendix}
%\label{sec:appendix}
%
%%%%%%%%%%%%%%%%%%%%%%%%%%%%%%%%%%%%%%%%%%%%%%%%%
%\appendix
%\end{appendix}
%%%%%%%%%%%%%%%%%%%%%%%%%%%%%%%%%%%%%%%%%%%%%%%%%
%
% References
%
%%%%%%%%%%%%%%%%%%%%%%%%%%%%%%%%%%%%%%%%%%%%%%%%%
%\section*{references}

%\bibliography{../../REFERENCES/references}     
%\bibliographystyle{prsty}
%\end{thebibliography}
%%%%%%%%%%%%%%%%%%%%%%%%%%%%%%%%%%%%%%%%%%%%%%%%%
%
% \Figures
%
%%%%%%%%%%%%%%%%%%%%%%%%%%%%%%%%%%%%%%%%%%%%%%%%%
%\Figure{}

\end{document}